# CAFFEINE: Template-Free Symbolic Model Generation of Analog Circuits via Canonical Form Functions and Genetic Programming

Trent McConaghy, Tom Eeckelaert, Georges Gielen

K.U. Leuven, ESAT-MICAS
Kasteelpark Arenberg 10
B-3001 Leuven, Belgium

## Abstract

*This paper presents a method to automatically generate compact symbolic performance models of analog circuits with no prior specification of an equation template. The approach takes SPICE simulation data as input, which enables modeling of any nonlinear circuits and circuit characteristics. Genetic programming is applied as a means of traversing the space of possible symbolic expressions. A grammar is specially designed to constrain the search to a canonical form for functions. Novel evolutionary search operators are designed to exploit the structure of the grammar. The approach generates a set of symbolic models which collectively provide a tradeoff between error and model complexity. Experimental results show that the symbolic models generated are compact and easy to understand, making this an effective method for aiding understanding in analog design. The models also demonstrate better prediction quality than posynomials.*

## 1 Introduction

Symbolic models of analog circuits have many applications. Fundamentally, they increase a designer's understanding of a circuit, which leads to better decision-making in circuit sizing, layout, verification, and topology design. Automated approaches to symbolic model generation are therefore of great interest.

In *symbolic analysis*, models are derived via topology analysis. [1] is a survey. Its main weakness is that it is limited to linear and weakly nonlinear circuits.

Leveraging SPICE simulations in modeling is promising because simulators readily handle nonlinear circuits, as well as environmental effects, manufacturing effects, and different technologies. Simulation data has been used to train neural networks as in [2,3,4]. However, such models provide no insight.

The aim of *symbolic modeling* is to *use simulation data* to generate interpretable mathematical expressions that relate the circuit performances to the design variables.

In [5,6], symbolic models are built from a posynomial template. The main problem is that the models are constrained to a template, which restricts the functional form and in doing so also imposes bias. Also, the models have dozens of terms, limiting their interpretability. Finally, the approach assumes posynomials can fit the data; in analog circuits there is no guarantee of this, and one might never know in advance.

The problem we address in this paper is how to generate symbolic models with more open-ended functional forms (i.e. without a pre-defined template), for arbitrary nonlinear circuits, and at the same time ensure that the models are interpretable. A target flow that reflects these goals is shown in Figure 1.

We approach the question by using genetic programming (GP) [7] as a starting point. GP generates symbolic expressions without the using a template, but those functions are overly complex. So, we extend GP via a grammar specifically designed to have *interpretable* symbolic models. We name the approach CAFFEINE: Canonical Functional Form Expressions in Evolution.

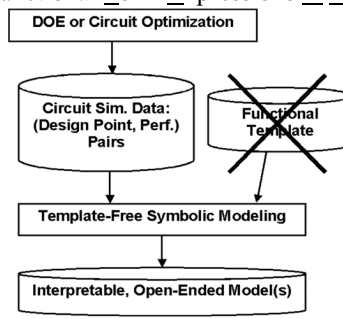

**Figure 1: Template-free symbolic modeling flow**

The contributions of this paper are as follows:
- To the best of our knowledge, a first-ever tool to do template-free symbolic modeling, with the flexibility of SPICE simulations therefore allowing modeling of any nonlinear circuits.
- A means to make the models compact and understandable, yet with arbitrary accuracy; in fact providing a tradeoff between accuracy and complexity. Final models are highly predictive.
- For GP, a specially designed grammar and related operators to ensure that all functions explored follow a canonical form, making them directly interpretable.

This paper is organized as follows. Section 2 defines the problem. Section 3 gives background on GP, on which sections 4 and 5 build to describe CAFFEINE and the grammar. Section 6 has results; section 7 concludes.

## 2 Problem Formulation

The problem that we address is formulated as follows:
*Given:*
- A set of $\{\mathbf{x}(t),y(t)\}, t=1..N$ data samples where $\mathbf{x}(t)$ is a $d$-dimensional design point $t$ and $y(t)$ is a corresponding circuit performance value measured from simulation of that design.
- **No** model template

*Determine:*
- A set of symbolic models $f^* \in F$ that together provide the optimal tradeoff between prediction error and some measure of complexity.

Speed of model building is *not* considered a goal at this point; that is left to future research.

## 3 Background: Genetic Programming

Genetic Programming (GP) [7] is an evolutionary algorithm, with the distinguishing characteristic that GP individuals (points in the design space) are *trees*.

GP has issues to be addressed before it can be useful in symbolic model generation. GP-evolved functions can be notoriously *complex* and *un-interpretable*; e.g. [7] showed functions so bloated that they take up a full page of dense text. Also, *overfitting* is a risk because prediction quality does not influence model choice.

The functional form of results from canonical GP is completely unrestricted. While this sounds great compared to the restrictions of fixed-template regression, it actually goes a little too far. Most importantly, an unrestricted form is almost always difficult to analyze. Also, an unrestricted form can cause undesirable biases in the search, such as tuning too many parameters which may even be redundant, or making it difficult for to add / remove basis functions. The challenge is to find a way to restrict the form enough to overcome these problems, without constraining away any possible forms.

## 4 CAFFEINE

CAFFEINE uses GP as a starting point, but extends it in order to properly address template-free symbolic modeling. It attacks the issues of complexity and interpretability in two main ways: a multi-objective approach that provides a tradeoff between error and complexity, and a specially designed grammar to constrain the search to specific functional forms without cutting out good solutions. It also performs special post-processing to further improve models. In CAFFEINE, the overall expression is a linear sum of weighted basis functions; therefore, each individual is a *set* of GP trees.

### 4.1 Multi-Objective Approach

CAFFEINE uses a state of the art *multi-objective* evolutionary algorithm, namely NSGA-II [8]. NSGA-II returns a set of individuals that, collectively, trade off error and complexity (i.e. a nondominated set).

"Error" is normalized mean-squared error. "Complexity" is dependent on the number of basis functions, the number of nodes in each tree, and the exponents of "variable combos" (VCs, described later):

$$\text{complexity}(f) = \sum_{j=1}^{M_f} (w_b + \text{nnodes}(j) + \sum_{k=1}^{nvc(j)} \text{vccost}(vc_{k,j})) \quad (1)$$

where $w_b$ is a constant to give a minimum cost to each basis function, nnodes($j$) is the number of tree nodes of basis function $j$, nvc($j$) is number of VCs of basis function $j$, and $\text{vccost}(vc) = w_{vc} \sum_{\dim=1}^{d} \text{abs}(vc(\dim))$.

The approach accomplishes *simplification during generation* by maintaining evolutionary pressure towards lower complexity. The user avoids an *a priori* decision on error or complexity because the algorithm generates a set of models that provide tradeoffs of alternatives.

## 5 Grammar and Operators

In GP, a means of constraining search is via a grammar, as in [9]. Evolutionary operators must respect the derivation rules of the grammar, i.e. only subtrees with the same root can be crossed over, and random generation of trees must follow the derivation rules. A basis function is the leaf nodes (terminal symbols) of the tree; internal nodes (nonterminal symbols) reflect the underlying structure; the tree root is the start symbol.

Even though grammars can usefully constrain search, none have yet been carefully designed for functional forms. In designing such a grammar, it is important to allow all functional combinations (even if just in one canonical form). This includes an arbitrary number of products of expressions, and of sums of expressions. Any desired single-input, dual-input, etc should be allowed.

The CAFFEINE grammar maintains functions in a "canonical form" and meets those goals:

```
REPVC  => 'VC' | REPVC '*' REPOP | REPOP
REPOP  => REPOP '*' REPOP | 1OP '(' 'W' '+'
          REPADD ')' | 2OP '(' 2ARGS ')' | ... 3OP, 4OP
          etc
2ARGS  => 'W' '+' REPADD ',' MAYBEW | MAYBEW ','
          'W' '+' REPADD
MAYBEW => 'W' | 'W' '+' REPADD
REPADD => 'W' '*' REPVC | REPADD '+' REPADD
2OP    => 'DIVIDE' |'POW' | 'MAX' | ...
1OP    => 'INV' | 'LOG10' | ...
```

Terminal symbols are in quotes. Each nonterminal symbol has a set of derivation rules separated by '|'. The start symbol is REVPC; one tree is used for each basis function; basis functions are linearly weighted using least-squares learning. Basis function operators include: creating a new individual by randomly choosing >0 basis function from each of 2 parents; deleting a random basis function; adding a randomly generated tree as a basis function; copying a subtree from one individual to make a new basis function for another.

The root is a product of variables and/or nonlinear functions (REPVC and REPOP). Within each nonlinear function is a weighted sum of basis functions (REPADD). Each basis function can be, once again, a product of variables and/or nonlinear functions. And so on.

The grammar is context-free, with two exceptions for the sake of enhanced search:

- *Weights* (W). A real value is stored in the range $[-2*B, +2*B]$ at each W node. During interpretation of the tree the value is transformed into $[-1e+B, -1e-B] \cup [0.0] \cup [1e-B, 1e+B]$. B is user-set, e.g. 10. In this way parameters can take on very small or very large negative or positive values. Zero-mean Cauchy mutation [10] is an operator on the real value.
- Single-basis rational combinations of *variables* (VC). With each VC a vector is stored, with integer value per design variable as the variable's exponent. An example vector is [1,0,-2,1], which means $(x_1 * x_4)/(x_3)^2$. For interpretability, real-valued and fractional-valued exponents are not allowed. VC operators include: one point crossover, and randomly adding or subtracting to an exponent value.

POW(a,b) is $a^b$. Via 2ARGS with MAYBEW, either the base or the exponent (but not both) can be constants.

The designer can turn off any of the rules if they are considered unwanted or unneeded. For example, one could easily restrict the search to polynomials or rationals, or remove potentially difficult-to-interpret functions such as sin and cos. The designer could change or extend the operators or inputs, e.g. include $w_i$, $l_i$, and $w_i/l_i$.

## 5.1 CAFFEINE Post-Processing

After the evolutionary run is complete, *simplification after generation* (SAG) is performed on each of the final set of models in the tradeoff. SAG is accomplished via the Predicted Residual Sums of Squares (PRESS) statistic $\xi^{(-1)}(t)$ [11] coupled with forward regression [12]. PRESS approximates leave-one-out cross-validation on the linear parameters; forward regression prunes basis functions that harm predictive ability. This gives predictive robustness to the linear parameters.

After that, the tradeoff models are evaluated on test data, and filtered down to only models that are on the tradeoff of *testing* error and complexity. Such a final step might not be possible with more deterministic approaches having more homogenous results, but the stochastic nature of CAFFEINE, causing more heterogeneous results, makes such filtering possible.

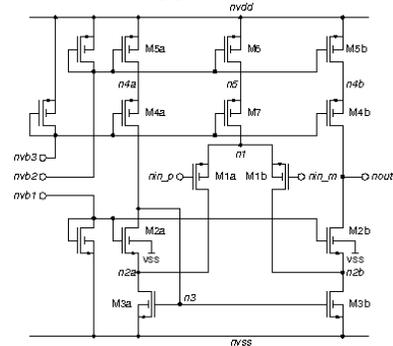

**Figure 2: Schematic of high-speed CMOS OTA**

## 6 Experiments

### 6.1 Experimental Setup

A prototype CAFFEINE system was written in about 2000 lines of Matlab code. The grammar was defined in a separate text file and parsed by the CAFFEINE system.

Single-input operators allowed were: $\sqrt{x}$, $\log_e(x)$, $\log_{10}(x)$, $1/x$, $abs(x)$, $x^2$, $\sin(x)$, $\cos(x)$, $\tan(x)$, $\max(0, x)$, and $\min(0,x)$, $2^x$, $10^x$, where $x$ is an expression. Double-input operators allowed are $x_1+x_2$, $x_1*x_2$, $\max(x_1,x_2)$, $\min(x_1,x_2)$, $power(x_1,x_2)$, and $x_1/x_2$. Also, lte(*testExpr*, *condExpr*, *exprIfTestLessThanCond*, *elseExpr*) and lte(*testExpr*, 0, *exprIfTestLessThan0*, *elseExpr*) were used. Any input variable could have an exponent in the range $\{…,-2, -1, 1, 2, …\}$. While real-valued exponents could have been used, that would have harmed interpretability.

The circuit being modeled is a high-voltage CMOS OTA as shown in Figure 2. The goal is to discover expressions for low-frequency gain ($A_{LF}$), unity-gain frequency ($f_u$), phase margin (*PM*), input-referred offset voltage ($v_{offset}$), and the positive and negative slew rate ($SR_p$, $SR_n$). To allow a direct comparison to the posynomial approach [6], an almost-identical problem setup was used, as well as identical simulation data. The only difference is that because scaling makes the model less interpretable, neither the inputs nor outputs were scaled; the one exception is that $f_u$ is log-scaled so that mean-squared error calculations and linear learning are not wrongly biased towards high-magnitude samples of $f_u$.

The technology is 0.7 $\mu m$ CMOS. The supply voltage is 5V. $V_{th,nom}$ is 0.76V and –0.75V for NMOS and PMOS devices, respectively. The load capacitance is 10 pF.

Good training data is essential to the methodology. The choice of design variables and sampling methodology determines the extent to which the designer can make inferences about the physical basis, and what regions of design space the model is valid in. We used an operating-point driven formulation [13], where currents and transistor drive voltages comprise design variables (13 variables in our case). Device sizings could have been used instead; it all depends on designer preference. Full orthogonal-hypercube Design-Of-Experiments (DOE) sampling of design points was used, with scaled dx=0.1 (the simpler problem of dx=0.01 is ignored in this paper) to have 243 samples with three simulations each, some of which did not converge. Simulation time for one sample was about 1 s, or 4 min for all samples; this is fully dependent on the circuit, analyses, and experimental design being used. These samples, otherwise unfiltered, were used as training data inputs. Testing data inputs were also sampled with full orthogonal-hypercube DOE and 243 samples, but with dx=0.03. Thus, in this experiment we are creating a somewhat localized model; one could just as readily model a broader design space.

The run settings were: maximum number of basis functions = 15, population size 200, 5000 generations, maximum tree depth 8, and parameter range $[-1e+10, -1e-10] \cup [0.0] \cup [1e-10, 1e+10]$. All operators had equal probability, except parameter mutation was 5x more likely. Complexity measure settings were $w_b = 10$, $w_{vc} = 0.25$. Just one run was done for each performance goal, for 6 runs total. (The aim was proof-of-concept, not efficiency.) Each run took about 12 hours on a 3 GHz Pentium IV Linux workstation. After each run, SAG (section 5.1) was done, taking about 10 min.

We use normalized mean-squared error on the training data and separate testing data, which are standard measurements in regression literature. Testing error is ultimately the more important measure. These measures are identical to two of the three posynomial "quality of fit" measures [6]: $q_{wc}$ is training error, and $q_{tc}$ is testing error. ($q_{wc}$ and $q_{tc}$ are identical as long as the constant 'c' in the denominator is zero, which [6] did.) We ignore $q_{oc}$, which measured the error at just one training point.

## 6.2 Results and Discussion

Let us first see if CAFFEINE generates tradeoffs between training error ($q_{wc}$) and complexity, as expected. Figure 3 illustrates tradeoff results. In each instance, CAFFEINE generates a tradeoff of about 50 different models. As expected, a zero-complexity model (i.e. just a constant) has the highest training error of 10-25%; the highest complexities have the lowest training error, of 1-3%. Since only one run was done for each performance characteristic, the reliability of the algorithm is promising.

As expected, the number of basis functions usually rises with the complexity. This is not always the case, however, as larger trees increase complexity too; the plateaus and dips in the basis function curves show that this does indeed occur. In every case, CAFFEINE used the maximum allowed number of basis functions (15) to achieve the lowest error. Undoubtedly, error could have been reduced further, but models with 15 basis function models are already at the edge of interpretability.

The testing error ($q_{tc}$) is also shown in Figure 3. We see that unlike training error, it is not monotonically decreasing as complexity rises. This means that some less complex models are more predictive than more complex ones. However, if we prune the models down to the ones that give a tradeoff between testing error and complexity, we get the rightmost column of Figure 3. These 5-10 models for each performance goal are of the most interest.

It is notable that the testing error is lower than the training error in almost all cases. This sounds promising, but such behavior is rare in the regression literature, and made us question what was happening. It turns out there is a valid reason: recall that the training data is from extreme points of the sampling hypercube (scaled dx=0.10), and the testing data is internal to the hypercube (dx=0.03). This testing data tests *interpolation* ability. Thus, models that really *are* predictive should be able to interpolate well, even at the cost of a perfect fit to the extreme points. In any case, to validly have testing error lower than training error demonstrates the strength of the CAFFEINE approach.

Let us now examine the actual symbolic models generated by CAFFEINE. We ask: "what are all the symbolic models that provide less than 10% error in both training and testing data?" Table I shows those functions ($f_u$ has been converted to its true form by putting the generated function to the power of 10). We see that each form has up to 4 basis functions, not including the constant. For $v_{offset}$, a constant was sufficient to keep the error within 10%. We see that a rational functional form was favored heavily; at these target errors only one nonlinear function, ln( ), appears (for $A_{LF}$). That expression effectively says that the *order of magnitude* of some input variables is useful.

One can examine the equations in more detail to gain an understanding of how design variables in the topology affect performance. For example, $A_{LF}$ is inversely proportional to $i_{d1}$, the current at the OTA's differential pair. Or, $SR_p$ is solely dependent on $i_{d1}$ and $i_{d2}$ and the ratio $i_{d1} / i_{d2}$. Or, within the design region sampled, the nonlinear coupling among the design variables is quite weak, typically only as ratios for variables of the same transistor. Or that each expression only contains a (sometimes small) subset of design variables. Or, that transistor pairs M1 and M2 are the only devices affecting five of the six performances (within 10%).

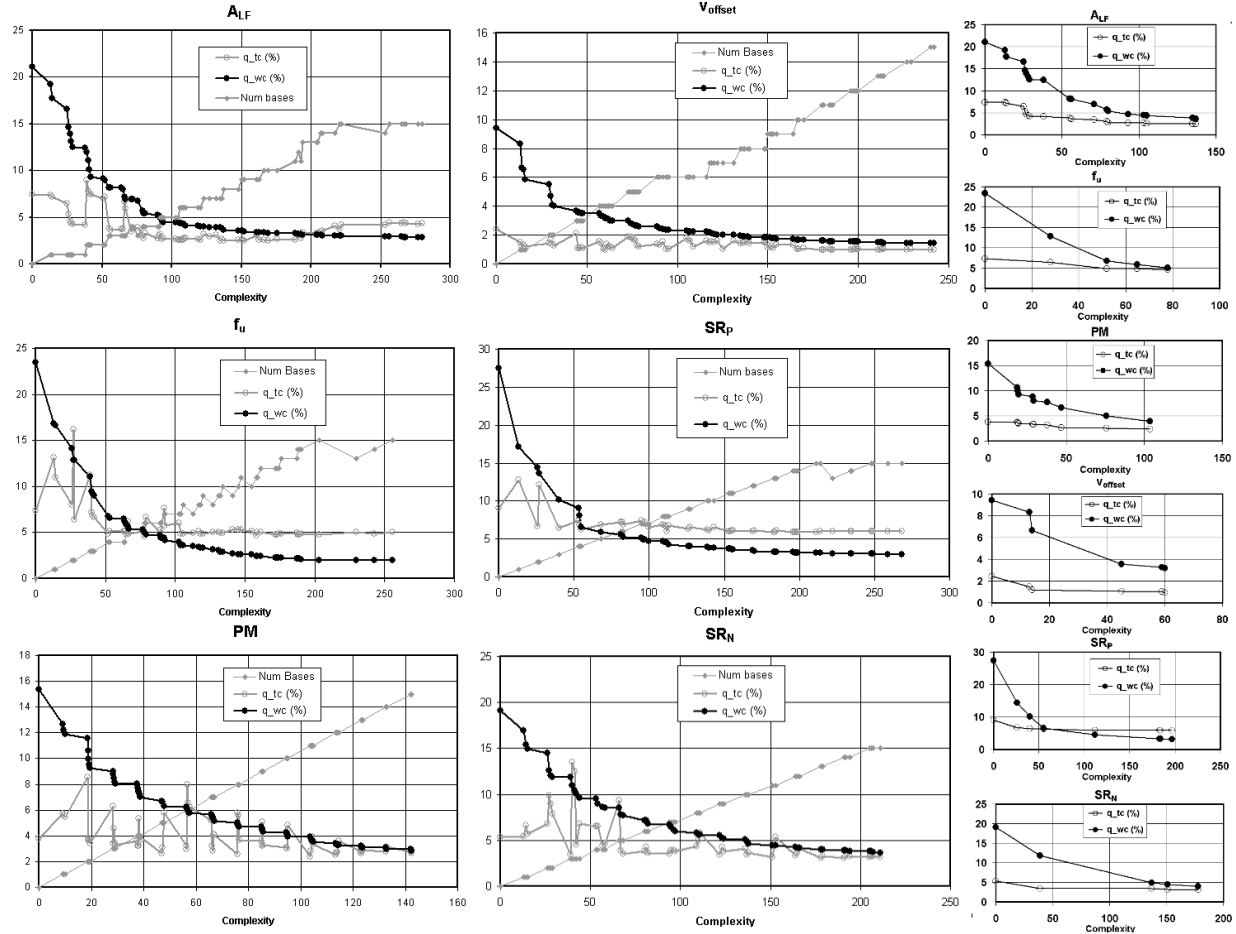

**Figure 3:** The two leftmost columns show generated models' training error ($q_{wc}$), testing error ($q_{tc}$), and number of bases vs. complexity for each performance goal; all models in the tradeoff of *training* error vs. complexity are shown. The rightmost column shows only models that are on the tradeoff of *testing* error vs. complexity too.

| Perf. | Target (%) | | Expression |
|---|---|---|---|
| | $q_{wc}$ | $q_{tc}$ | |
| $A_{LF}$ | 10 | 10 | -10.3 + 7.08e-5 / id1 + 1.87 * ln( -1.95e+9 + 1.00e+10 / (vsg1*vsg3)+ 1.42e+9 *(vds2*vsd5) / (vsg1*vgs2*vsg5*id2)) |
| $f_u$ | 10 | 10 | 10^( 5.68 - 0.03 * vsg1 / vds2 - 55.43 * id1+ 5.63e-6 / id1 ) |
| PM | 10 | 10 | 90.5 + 190.6 * id1 / vsg1 + 22.2 * id2 / vds2 |
| $v_{offset}$ | 10 | 10 | - 2.00e-3 |
| $SR_p$ | 10 | 10 | 2.36e+7 + 1.95e+4 * id2 / id1 - 104.69 / id2 + 2.15e+9 * id2 + 4.63e+8 * id1 |
| $SR_n$ | 10 | 10 | - 5.72e+7 - 2.50e+11 * (id1*id2) / vgs2 + 5.53e+6 * vds2 / vgs2 + 109.72 / id1 |

**Table I: CAFFEINE-generated symbolic models which have less than 10% training and testing error**

| Test error (%) | Train error (%) | PM Expression |
|---|---|---|
| 3.98 | 15.4 | 90.2 |
| 3.71 | 10.6 | 90.5 + 186.6 * id1 + 22.1 * id2 / vds2 |
| 3.68 | 10.0 | 90.5 + 190.6 * id1 / vsg1 + 22.2 * id2 / vds2 |
| 3.39 | 8.8 | 90.1 + 156.85 * id1 / vsg1 - 2.06e-03 * id2 / id1 + 0.04 * vgs2 / vds2 |
| 3.31 | 8.0 | 91.1 - 2.05e-3 * id2 / id1 + 145.8 * id1 + 0.04 * vgs2 / vds2 - 1.14 / vsg1 |
| 3.20 | 7.7 | 90.7 - 2.13e-3 * id2 / id1 + 144.2 * id1 + 0.04 * vgs2 / vds2 - 1.00 / (vsg1*vsg3) |
| 2.65 | 6.7 | 90.8 - 2.08e-3 * id2 / id1 + 136.2 * id1 + 0.04 * vgs2 / vds2 -1.14 / vsg1 + 0.04 * vsg3 / vsd5 |
| 2.41 | 3.9 | 91.1 - 5.91e-4 * (vsg1*id2) / id1 + 119.79 * id1 + 0.03 * vgs2 / vds2 - 0.78 / vsg1 + 0.03 * vsg1 / vsd5 -2.72e-7 / (vds2*vsd5*id1) + 7.11 * (vgs2*vsg4*id2) - 0.37 / vsg5 - 0.58 / vsg3 - 3.75e-6 * id2 - 5.52e-6 / id1 |

**Table II: CAFFEINE-generated models of PM, in order of decreasing error and increasing complexity**

By only putting the relevant variables into a model, the approach demonstrates the potential to provide expressions for circuits with significantly more variables.

One may improve understanding in another fashion: by examining expressions of varying complexity for a single performance characteristic. Low-complexity models will show the macro-effects; alterations to get improved error point show how the model is refined to handle second-order effects. Table II shows models generated for PM in decreasing training and testing error. A constant of 90.2, while giving 15 % training error, had only 4% test error. For better prediction, CAFFEINE injected two more basis functions; one basis being the current into the differential pair $i_{d1}$, the other basis, $i_{d2} / v_{ds2}$, the ratio of current to drain-source voltage at M2. The next model turns the input current term into a ratio $i_{d1} / v_{sg1}$. Interestingly, and reassuringly, almost all ratios use the same transistor in the numerator and denominator.

Such analyses achieve the aim of this tool: to improve understanding of the topology.

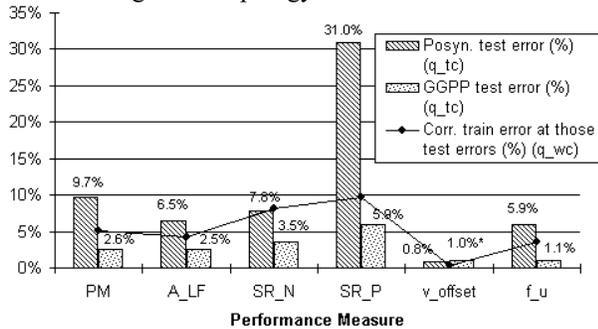

**Figure 4: Comparison of CAFFEINE testing error to posynomial testing error; also to training error**

We also compared CAFFEINE to the posynomial approach using the numbers in [5]. We first compare the test and training errors. To pick a model from a CAFFEINE-generated tradeoff for comparison, we fixed the training error to what the posynomial achieved, then compared testing errors. Results are in Figure 4. In one case, $v_{offset}$, CAFFEINE did not meet the posynomial training error (0.4%), so it probably could have for more basis functions; we instead picked an expression which very nearly matched the posynomial approach's testing error of 0.8%. What we saw in previous data, and we see again here, is that CAFFEINE has lower testing error than training error, which provides great confidence to the models. In contrast, in all cases but $v_{offset}$, the posynomials had higher testing error than training error, even on this interpolative data set. CAFFEINE models' testing errors were *2x to 5x* lower than the posynomial models. The exception is $v_{offset}$, where the posynomial achieves 0.8% testing error compared to 0.95% for CAFFEINE. With posynomials having weak prediction ability even in interpolation, in comparison to more compact models, one might question the trustworthiness of constraining models of analog circuits to posynomials.

## 7 Conclusion

This paper presented CAFFEINE, a tool which for the first time can generate interpretable, template-free symbolic models of nonlinear analog circuit performance characteristics. CAFFEINE is built upon genetic programming, but its key is a grammar that restricts symbolic models to a canonical functional form.

CAFFEINE generates a set of models that collectively trade off between error and complexity. Visual inspection of the models demonstrates that the models are interpretable. These models were also shown to be significantly better than posynomials in predicting unseen data.